\documentclass[iop]{emulateapj}

\usepackage{epsfig}
\usepackage{ulem}
\usepackage{amsmath}
\usepackage{subfigure}
\usepackage{natbib}
\usepackage{multirow}
\usepackage{rotating}
\usepackage{longtable}
\usepackage{txfonts}
\newcommand{\chemrev}{Chem.~Rev.}  %

\begin{document}

\title{Discovery of the ubiquitous cation NS$^+$ in space confirmed by laboratory spectroscopy}

\shorttitle{Discovery of NS$^+$}
\shortauthors{Cernicharo et al.}

\author
{
J. Cernicharo\altaffilmark{1,2},
B. Lefloch\altaffilmark{3}
M. Ag\'undez\altaffilmark{1,2},
S. Bailleux\altaffilmark{4},
L. Margul\`es\altaffilmark{4},
E. Roueff\altaffilmark{6},
R. Bachiller\altaffilmark{5},
N. Marcelino\altaffilmark{1,2},
B. Tercero\altaffilmark{1,5},
C. Vastel\altaffilmark{7},
E. Caux\altaffilmark{7}
}

\altaffiltext{1}{Group of Molecular Astrophysics. ICMM. CSIC. C/ Sor Juana In\'es de La Cruz 3, E-28049, Madrid, Spain}
\altaffiltext{2}{Dpt. of Molecular Astrophysics. IFF. CSIC. C/ Serrano 123, E-28006, Madrid, Spain}
\altaffiltext{3}{Universit\'e Grenoble Alpes, CNRS, IPAG, 38000 Grenoble, France}
\altaffiltext{4}{Laboratoire de Physique des Lasers, Atomes et Mol\'ecules, UMR 8523 CNRS, Universit\'e Lille, 59655 Villeneuve d'Ascq Cedex, France.}
\altaffiltext{5}{Observatorio Astron\'omico Nacional (OAN, IGN), Calle Alfonso XII, 3, 28014 Madrid, Spain}
\altaffiltext{6}{LERMA, Observatoire de Paris, PSL Research University, CNRS, Sorbonne UniversitŽs, UPMC Univ. Paris 06,  F-92190 Meudon, France}
\altaffiltext{7}{IRAP, Universit\'e de Toulouse, CNRS, UPS, CNES, Toulouse, France}

\date{Received December 14, 2017; accepted January 15, 2018}

\begin{abstract}
We report the detection in space of a new molecular species which has been characterized 
spectroscopically and fully identified
from astrophysical data. The observations were carried out with the 30m IRAM 
telescope\footnote{This work was based on observations carried out with the IRAM 30-meter 
telescope. IRAM is supported by INSU/CNRS (France), MPG (Germany) and IGN (Spain).}. The molecule 
is ubiquitous as its $J$=2$\rightarrow$1 transition has been found in cold molecular clouds, 
prestellar cores, and shocks. However, it is not found in the hot cores of Orion-KL and in 
the carbon-rich evolved star IRC+10216. 

Three rotational transitions in perfect harmonic 
relation $J'=2/3/5$ have been identified in the prestellar core B1b. 
The molecule has a $^1\Sigma$ electronic ground state and its $J$=2$\rightarrow$1 
transition presents the hyperfine structure characteristic
of a molecule containing a nucleus with spin 1. A careful 
analysis of possible carriers shows that the best candidate is NS$^+$. 
The derived rotational constant agrees within 0.3-0.7\% with ab initio calculations. 
NS$^+$ was also produced in the laboratory to unambiguously validate the astrophysical 
assignment. The observed rotational frequencies and determined 
molecular constants confirm the discovery of the nitrogen sulfide cation in space. 

The chemistry of NS$^+$ and related nitrogen-bearing species has been analyzed 
by means of a time-dependent gas phase model. The model reproduces well the observed 
NS/NS$^+$ abundance ratio, in the range 30-50, and indicates that NS$^+$ is formed 
by reactions of the neutral atoms N and S with the cations SH$^+$ and NH$^+$, respectively.
\end{abstract}

\keywords{ISM: clouds --- line: identification --- molecular data --- radio lines: ISM}

\section{Introduction}

In molecular clouds nitrogen is mainly in the form of N$_2$, which is unobservable 
directly via its rotational spectrum. Diazenylium (N$_2$H$^+$) is usually used as 
a tracer of N$_2$ \citep{Daniel2006,Daniel2007,Daniel2013}. 
Major N-bearing molecules in the gas phase of interstellar clouds are HCN, HNC, 
CN, NH$_3$, ammonium (NH$_4^+$), the protonated ion of NH$_3$ 
\citep[see][and references therein]{Cernicharo2013},
and NCCN/NCCNH$^+$ \citep{Agundez2015}.
Many other N-bearing species,
such as the radicals C$_n$N and the cyanopolyynes, have been studied in detail 
and their chemistry is rather well
understood \citep[see][and references therein]{Agundez2013}. However, only 
three N-bearing species containing sulfur have been detected so far: NS, 
HNCS, and HSCN. Nitrogen sulfide (NS) was detected in the early years of 
radioastronomy by \citet{Gottlieb1975} and it has been observed in practically 
all interstellar environments, even in comets \citep{Irvine2000}.
Although a recent study of the chemistry of sulfur in cold dense clouds 
has been carried out by \citet{Fuente2016},
the formation pathways of these molecules are far from well understood. 
The 
gas-phase chemistry of cold dark clouds is mainly based on the reactions 
between ions and neutrals \citep{Agundez2013}. However, only a small percentage ($\approx$15\%) 
of the detected species are ions (cations or anions). The last diatomic 
cation identified in space is NO$^+$, which was detected towards B1b 
with an abundance ratio NO/NO$^+$ $\sim500$ \citep{Cernicharo2014}, 
similar to the abundance ratio SO/SO$^+$ ($\sim1000$) observed in the 
same source by \citet{Fuente2016}. 

In this Letter we report on the discovery for the first time in 
space of the nitrogen sulfide cation, NS$^+$. This species has been 
fully characterized through three rotational transitions, one of 
them exhibiting hyperfine structure due to the presence of nitrogen, 
and its detection in space has been confirmed by laboratory spectroscopy. 

\section{Observations}

The observations presented in this paper are part of a complete spectral line survey 
at 3, 2, and 1 mm of a set of interstellar
clouds harboring different physical conditions within the IRAM large program 
ASAI\footnote{http://www.oan.es/asai} \citep{Lefloch2017}. They have been
complemented in this work with data at 3 mm of B1b and 
TMC1 \citep{Marcelino2005,Marcelino2007,Marcelino2009,Nuria2007,Cernicharo2012b,Cernicharo2013} 
and various dark clouds observed in a search for molecular anions \citep{Agundez2015b}.
 
All observations were performed between 2006 and 2017 with the IRAM 30m radiotelescope, 
located at Pico Veleta (Spain), using different sets of receivers and spectrometers 
(see references above). System temperatures were in the range 80-300 K. Pointing errors 
were always within 3''. 
Spectra were calibrated in antenna temperature corrected for 
atmospheric
attenuation using the ATM package \citep{Cernicharo1985,Pardo2001}. 
All data were 
processed using the GILDAS software package\footnote{\texttt{http://www.iram.fr/IRAMFR/GILDAS/}}.

\begin{figure*}
\begin{center}
\includegraphics[angle=0,scale=0.71]{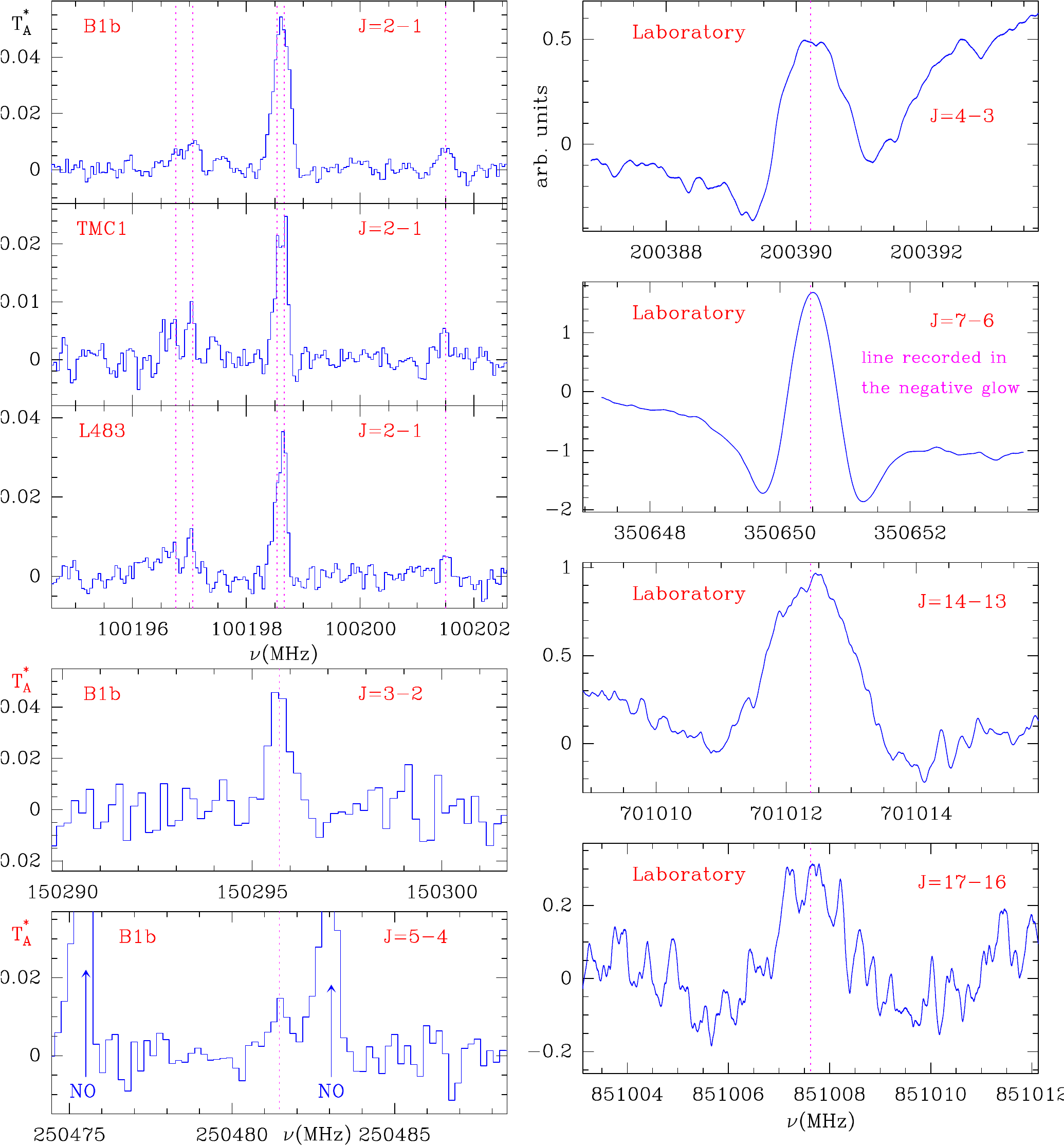}
\caption{\textbf{Left upper panels:} Observed feature at 100198.55 MHz in B1b, TMC1, and L483.
Vertical dashed lines indicate the position of the hyperfine
component calculated from the fit to this line and those
shown in the bottom panels. 
The hyperfine structure is computed for a nuclear spin of 1.
\textbf{Left bottom panels:} Observed lines in B1b corresponding to the $J$=3$\rightarrow$2
($\nu_0$=150295.611$\pm$0.080 MHz)
and $J$=5$\rightarrow$4 ($\nu_0$=250481.477$\pm$0.080 MHz)
transitions of NS$^+$.
Vertical dashed lines indicate the position of the calculated
frequencies. 
\textbf{Right panels:} 
Some laboratory transitions recorded in the positive column discharge
(except otherwise mentionned) in Ar, CS$_2$ and atmospheric air.
}
\label{Fig1}
\end{center}
\end{figure*}

\begin{table*}
\begin{center}
\caption[]{Source parameters and derived column densities for NS$^+$ and NS}
\begin{tabular}{|l | r| r| r| r| r| r| r| r| r| r| r| r|}
\hline
Source & $\alpha_{2000}$ & $\delta_{2000}$ & $V_{LSR}$      & $\Delta$v$^a$   &T$_A^*$$^b$ &$n$(H$_2$)    & $N$(H$_2$) & $T_k$ & $N$(NS$^+$)& $N$(NS) & $R^c$& Ref.\\
       &  (h:m:s)        & ($^o$ $'$ $''$) & (km\,s$^{-1}$) &(km\,s$^{-1}$)& (mK) &(cm$^{-3}$)   & (cm$^{-2}$) & (K)    & (cm$^{-2}$) & (cm$^{-2}$) & &\\ 
       &                 &                 &                & & &$\times\,10^5$& $\times\,10^{22}$&     &$\times\,10^{10}$&$\times\,10^{12}$ & &\\ 
\hline
IRAS4A           & 03:29:10.42 & 31:13:32.2& 7.4 & 1.19 &  17.0&20.0 & 37.0 &  30   & 22.6& 10.0& 44 & 1\\
B1b              & 03:33:20.80 & 31:07:34.0& 6.5 & 0.97 &  54.5& 1.0 & 15.0 &  12   & 20.0& 10.0& 50 & 2\\
L1495B           & 04:15:41.80 & 28:47:46.0& 7.7 & 0.42 &  12.6& 0.1 &  1.2 &  10   &  4.0&  ...& ...& 3\\
L1521F           & 04:28:39.80 & 26:51:35.0& 6.4 & 0.38 &  13.7&11.0 & 14.0 &  10   &  1.7&  ...& ...& 3\\
TMC1             & 04:41:41.88 & 25:41:27.0& 5.7 & 0.69 &  23.8& 0.5 &  1.0 &  10   &  5.2&  1.7& 33 & 2\\
L1544            & 05:04:17.21 & 25:10:42.8& 6.9 & 0.59 &  11.4& 0.8 &  9.4 &  12   &  2.3&  1.2& 52 & 4\\
Lupus-1A         & 15:42:52.40 &-34:07:53.5& 5.0 & 0.47 &  15.8& 5.0 &  1.5 &  14   &  4.2&  ...& ...& 3\\
L483             & 18:17:29.80 &-04:39:38.0& 5.3 & 0.70 &  36.7& 0.4 &  3.0 &  10   &  8.5&  ...& ...& 3\\
Serpens South 1a & 18:29:57.90 &-01:56:19.0& 7.4 & 1.03 &  23.0& 0.1 &  2.0 &  11   &  8.5&  ...& ...& 3\\
L1157-mm         & 20:39:06.30 & 68:02:15.8& 2.6 & 1.20 &  12.0& 0.6 &  0.6 &  12   &  9.1&  2.6& 29 & 5\\
L1157-B1         & 20:39:10.20 & 68:01:10.5& 2.6 & 2.95 &   8.0& 5.0 &  0.1 &  60   & 40.0& 13.0& 33 & 5\\
Orion KL (HC$^d$)& 05:35:14.50 &-05:22:30.0& 5.0 & ...  & ...  &500.0& 42.0 & 225   & $\leq$100 &3000& $\geq$3000 & 6\\
Orion KL (CR$^e$)& 05:35:14.30 &-05:22:37.0& 7.5 & ...  & ...  & 10.0&  7.5 & 110   & $\leq$100 & 500& $\geq$ 500 & 6\\
SgrB2 (envelope) & 17:47:43.00 &-28:23:12.1&66.0 & 10.0 &  35.0& 0.1 &  ... &  30   &       100 & 50 & 50 & 7\\ 
\hline
\end{tabular} 
\end{center}
{
$^{\rm a,b}$ Observed $\Delta$v and T$_A^*$ for the $J=2 \rightarrow 1$ main component of NS$^+$.\\
$^{\rm c}$ $R$ is the NS/NS$^+$ abundance ratio.\\
$^{\rm d,e}$ The hot core and compact ridge components of the cloud.\\ 
References for adopted densities and temperatures: (1) \citet{Lefloch1998}; 
(2) \citet{Cernicharo2012b,Cernicharo2013,Fuente2016,Marcelino2007}; (3) \citet{Agundez2015b,Loison2016};
(4) \citet{Quenard2017}; (5) \citet{Lefloch2012}; (6) \citet{Tercero2010}; (7) Tercero et al., in preparation.
}
\label{Table1}
\end{table*}

\section{Results}

In all observed sources, a line at 100198.55$\pm$0.08 MHz emerges as 
one of the strongest unidentified features. The frequency
uncertainty includes a source velocity error of
0.15 km\,s$^{-1}$.
Figure 1 shows the observed spectra in TMC1, B1b, and L483, and
clearly suggests the presence
of hyperfine structure of a low-$J$ transition, 
most likely the $J=2\rightarrow1$ transition of a molecule containing a nucleus 
with spin 1. Two additional lines at frequencies 3/2 (150295.611$\pm$0.08 MHz) 
and 5/2 (250481.477$\pm$0.08 MHz) have been found in B1b (see Figure 1 
bottom panels). They correspond to the $J=3\rightarrow2$ and $5\rightarrow4$ 
transitions of a new molecular species. None of these lines can be identified 
using the catalogues CDMS \citep{Muller2005} and JPL \citep{Pickett1998}. 
The possible candidates are discussed in next section.

The three unidentified lines, including the hyperfine structure of the 
$J=2\rightarrow1$ one, have been fitted to the standard Hamiltonian of 
a linear molecule including nuclear quadrupolar coupling
providing $B=25049.919\pm0.010$ MHz, $D=35.42\pm0.20$ kHz, 
and $eQq=-5.93\pm0.10$ MHz (for a nuclear spin of 1).

The $J=2\rightarrow1$ line has been detected towards all cold sources that 
we have examined, but also in shocks, and in the warm envelopes of SgrB2 
and the hot corino IRAS4A in which the kinetic temperature is $\simeq$30 K 
(see Figure 2). However, it is not detected towards the hot cores of 
Orion KL (data from Tercero et al. 2010), SgrB2(N) 
(public data from Belloche et al. 2013) and the carbon-rich star 
IRC+10216. Physical parameters for the sources are given in Table 1.

\subsection{Astrophysical identification of NS$^+$}

We have checked that 2/3/5 harmonic relation found in B1b cannot 
correspond to higher values of the rotational quantum number by searching 
for lines that could be present if $B$ was half or a third of the measured 
value (i.e., harmonic relations 4/6/10 or 6/9/15). We conclude that this 
harmonic relation is the only possible one characterizing spectroscopically 
the three observed lines. We have also checked that the lines are not associated 
to any other nearby feature of similar intensity ($\pm$1 GHz) that could indicate 
a doublet electronic state. We conclude that the molecule is linear
with a $^1\Sigma$ electronic ground state. 

\begin{figure}
\begin{center}
\includegraphics[angle=0,scale=0.70]{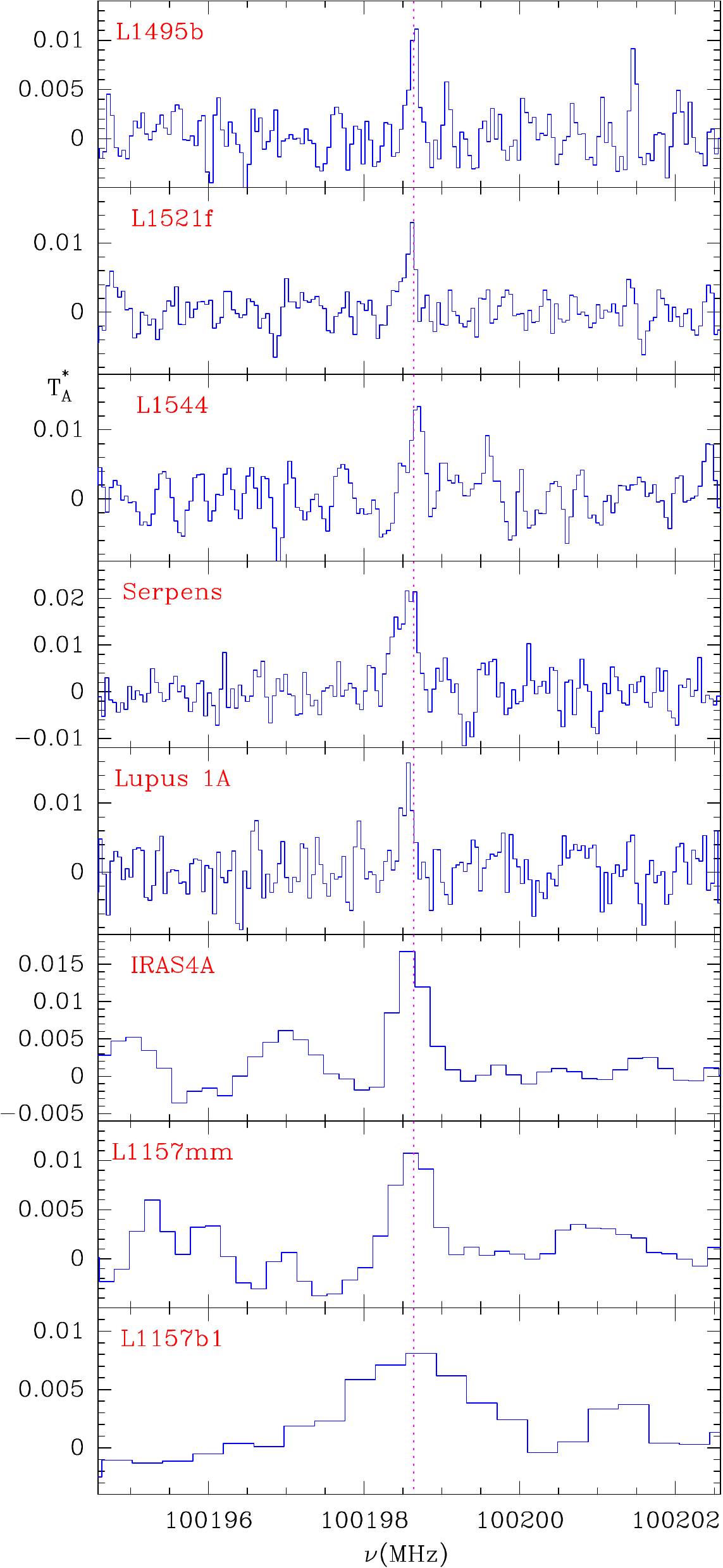}
\caption{Observed $J$=2$\rightarrow$1 transition of NS$^+$ towards
different sources.}
\label{Fig2}
\end{center}
\end{figure}

Being a close-shell molecule with $B\sim25.05$ GHz), 
the number of candidates is not very large.
The species has to include only H, N, C, O or S (silicon- or metal-bearing species 
are not detected in cold dark clouds). Moreover, the molecule has to be diatomic, 
or triatomic containing H, in order to have this relatively large rotational 
constant. Furthermore,
the molecule has to contain N as it is the only nucleus with spin 1. Hence, our 
carrier is NX or HXN/HNX (X=C,S,O), neutral or cation. Anions are unlikely  
because negative ions have not been detected in most of the 
observed sources. HCN and HNC have a much larger rotational constant. The same 
applies to HNO. The best candidates could be HCS, HSC, HNS, HSN (neutral or cations). 
HCS and HSC are well studied in the laboratory, are slightly asymmetric,
have doublet electronic ground states and their (B+C)/2 values are around 20 GHz. 
Their cations could have a slightly larger rotational constant but for HCS$^+$ 
$B\sim21.4$ GHz \citep{Margules2003}. No laboratory data are available for HSC$^+$. 
$Ab$ $initio$ calculations by \citet{Puzzarini2005} indicated that it is 
bent with (B+C)/2 too low.
HSN and HNS are both bent with singlet and triplet ground electronic states,
respectively \citep{Yaghlane2014}. Their cations are also bent with doublet ground
electronic states and too low (B+C)/2 \citep{Yaghlane2014,Trabelsi2016}.
Hence, no reasonable triatomic species can be found as carrier of
our lines.

If the carrier is a diatomic species it must contain sulfur or phosphorus in order 
to have the correct order of magnitude for the rotational constant. None of the well 
known diatomic species (CS, SO, PS, PO, PN, NS, CP) fit the observed $B$. All their 
cations, except NS$^+$ and PO$^+$, have spin multiplicity $>$1 in their ground electronic 
state. For example, CS$^+$, has a $^2\Sigma^+$ electronic state, $B=25.9$ GHz and 
$D=41.3$ kHz \citep{Bailleux2008}, while CP$^+$ has a $^3\Pi$ ground electronic state 
with $B_e\sim21.5$ GHz \citep{Bruna1980}. PO$^+$ has a singlet electronic state with 
$B=23.512$ GHz \citep{Petrmichl1991}. Hence, the only remaining plausible candidate 
is NS$^+$.

Several $ab$ $initio$ calculations have been performed for NS$^+$ 
\citep{Dyke1977,Karna1986,Peterson1988,Yaghlane2005,Yaghlane2009}. 
All of them predict a $^1\Sigma^+$ ground electronic state. From 
the photoelectron spectrum of NS, $r_e$(NS$^+$) was estimated to 
be 1.440$\pm$0.005 $\AA$ \citep{Dyke1977}. The calculated values for 
$\alpha_e$, the vibrational contribution to the rotational constant, 
range (in MHz) between 120 \citep{Yaghlane2009} and 180
\citep{Peterson1988}. These estimations provide an equilibrium rotational 
constant 24.9-25.2 GHz and $B_0$ between 24.84 and 25.11 GHz. The agreement 
with our rotational constant, 25.05 GHz, is within 0.3-0.7\%. 
The estimated distortion constant by \citet{Peterson1988}, 36 kHz, is 
also in excellent agreement with the value we have
derived from our lines (35.4 kHz). Moreover, \citet{Peterson1988} 
computed the quadrupole coupling constant, $eQq$, to be -6.54 MHz, which compares 
very well with our value of -5.95 MHz. Consequently, the excellent agreement between 
our rotational constants and the $ab$ $initio$ calculations strongly support NS$^+$ 
as the carrier of our lines which
has been finally confirmed in the laboratory (see section 3.3).

\subsection{Column densities}

B1b is the only source where we have more than one NS$^+$ line and thus we can 
study its excitation conditions. The calculated $ab$ $initio$ dipole moment of 
NS$^+$ is 2.17 D \citep{Peterson1988}, a value which requires
a large density to pump efficiently  the $J$=5 rotational level. No collisional 
rates are available for NS$^+$. The molecule is isoelectronic to CS and thus we 
have adopted the collisional rates for CS calculated by \citet{Lique2007} scaled up by a factor 5
to take into account that NS$^+$ is an ion (see, e.g., the case of HCO$^+$ and CO).

The molecule has been implemented into the MADEX code \citep{Cernicharo2012} which 
provides standard LTE or LVG calculations based on the formalism of
\citet{Goldreich1974}. 
For densities $n$(H$_2$) $>$ 10$^6$ cm$^{-3}$ 
the rotational levels up to $J$ = 5 could be nearly thermalized. For typical values 
of $n$(H$_2$) in cold dense cores of a few 10$^5$ cm$^{-3}$, $T_k$ = 10-15 K, the 
rotational temperature of the $J$ = 2$\rightarrow$1 line will be just 2-3 K below $T_k$. 
However, for lower densities the $J$ = 2$\rightarrow$1 transition could be subthermally 
excited. Hence, we have estimated the column density of NS$^+$ using the LVG approximation 
for all observed clouds. Although the collisional rates are very uncertain 
our calculations indicate that the J=2-1 line is close to thermalization
for the physical conditions given in Table 1.

The three lines observed towards B1b, assuming that B1b fills the beam at the 
three observed frequencies, can be fitted with $n$(H$_2$) = 10$^5$ cm$^{-3}$, 
$T_k$ = 12 K \citep{Marcelino2007,Cernicharo2012b}, and $N$(NS$^+$)=2$\times$10$^{11}$
cm$^{-2}$. The beam averaged total gas column density towards B1b could be a few 
10$^{22}$ cm$^{-2}$ \citep{Cernicharo2012b}, and thus the fractional abundance of 
NS$^+$ in B1b is a few 10$^{-12}$ relative to H$_2$. 

We have derived column densities of NS in the sources for which data is 
available assuming LTE because no collisional 
rates are available for this species. For TMC1, L1544 and the cold envelope 
of SgrB2 only data on the $J$=5/2$\rightarrow$3/2 are available. However,
for For B1b, IRAS4a, L1157mm,
and L1157B1 several transitions have been observed (from 
$J$=5/2$\rightarrow$3/2 up to 11/2$\rightarrow$9/2).
No NS data was available 
for the other sources. LTE could underestimate the column density because
the excitation temperature of the high $J$ lines will be subthermal. In order to 
evaluate the extent of non-LTE excitation we have adopted the collisional rates 
of CS for NS and computed its excitation conditions neglecting its fine and hyperfine 
structure. 
The excitation temperature obtained for the unsplitted lines was assumed to be 
the same for all fine and hyperfine components. For the physical conditions of 
the clouds, the LVG method results in NS column densities 1.5-2.5 times larger 
than assuming LTE. Hence, the final NS column densities are uncertain
within a factor 2.

The derived column densities of NS$^+$ (assuming $E(J=2)$=5.01340\,cm$^{-1}$)
and NS are given in Table 1. The NS/NS$^+$ 
abundance ratio ranges between 30 and 50. In B1b this abundance ratio is 50, which 
is $\sim10$ and $\sim20$ times smaller than the abundance ratios NO/NO$^+$ \citep{Cernicharo2014} 
and SO/SO$^+$ \citep{Fuente2016}, respectively. 

\begin{figure*}
\begin{center}
\includegraphics[angle=0,width=0.98\textwidth]{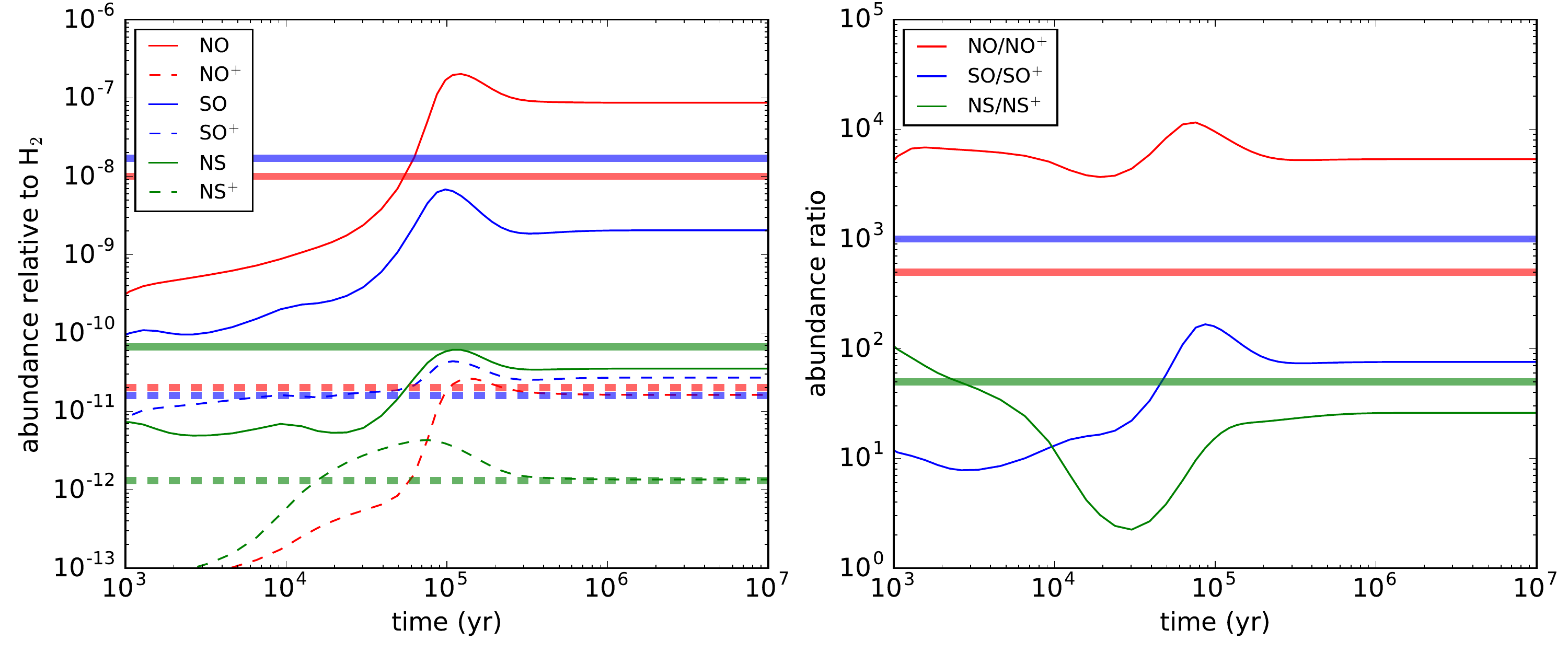}
\caption{Calculated abundances of NO, NO$^+$, SO, SO$^+$, 
NS, and NS$^+$ relative to n(H$_2$) as a function of time for 
n(H$_2$)=10$^5$ cm$^{-3}$  and T = 12K (left) and 
the corresponding ratios  NO/NO$^+$, SO/SO$^+$, and NS/NS$^+$ (right). 
The observed values in B1b are indicated by horizontal thick lines.}
\label{Fig_chemistry}
\end{center}
\end{figure*}

\subsection{Laboratory detection}
To corroborate the astrophysical discovery, we set up an experiment to detect NS$^+$ by
rotational spectroscopy below 1~THz. The salient features of the spectrometer 
have been thoroughly described in \citet{Bailleux17}.
Briefly, millimeterwave radiation was obtained by the frequency multiplication of an E8257D synthesizer
referenced to a GPS-locked Rubidium atomic clock. 
The instrumental resolution of the frequency multiplier chain is better than 1 Hz at 1 THz
and the achievable resolution is typically 500 kHz at 0.5 THz.
The cation was produced in a Pyrex absorption cell by glow discharging a mixture of N$_2$ and
CS$_2$ (Merck).
H$_2$S (purity: 99.9999\%) and OCS could also used as sources of sulfur.
A solenoid coil wound on the cell could be used to magnetically extend the negative glow region.


\begin{table}
\caption[]{Laboratory transition frequencies (in MHz) of NS$^+$ below 1 THz}
\vspace{-0.4cm}
\begin{center}
\begin{tabular}{c  c  c  c  c  c  }
\hline
$J' \leftarrow J''$ & $\nu_{obs}$ & $\Delta\nu^{a}$ &$J' \leftarrow J''$ & $\nu_{obs}$ & $\Delta\nu^{\rm a}$ \\
\hline
 $3\leftarrow2$ & 150295.607 & $ 0.002$   & $12\leftarrow11$ & 600955.245 & $-0.004$ \\
 $4\leftarrow3$ & 200390.216 & $ 0.003$   & $13\leftarrow12$ & 650989.258 & $-0.021$ \\
 $5\leftarrow4$ & 250481.463 & $ 0.007$   & $14\leftarrow13$ & 701012.396 & $ 0.025$ \\
 $6\leftarrow5$ & 300568.504 & $ 0.011$   & $15\leftarrow14$ & 751023.687$^{b}$ & $0.013^{c}$ \\
 $7\leftarrow6$ & 350650.500 & $ 0.019$   & $16\leftarrow15$ & 801022.380$^{b}$ & $0.016^{c}$ \\
 $8\leftarrow7$ & 400726.582 & $ 0.003$   & $17\leftarrow16$ & $851007.624$ & $0.013$ \\
 $9\leftarrow8$ & 450795.933 & $-0.014$   & $18\leftarrow17$ & 900978.540$^{b}$ & $0.027^{c}$\\
 $10\leftarrow9$ & 500857.724 & $-0.018$   & $19\leftarrow18$ & 950934.324$^{b}$ & $0.033^{c}$ \\
 $11\leftarrow10$ & 550911.123$^{b}$ & $ 0.006^{c}$ & $20\leftarrow19$ & 1000874.121$^{b}$ & $0.040^{c}$ \\
\hline
\end{tabular}
\end{center}
\vspace{-0.2cm}
{
$^{a}\Delta\nu = \nu_{obs} - \nu_{calc}$.\\
$^{b}$Predicted transition.\\
$^{c}$Uncertainty is given instead of $\Delta\nu$.\\
Derived molecular constants: $B_0$= 25049.89843(50) MHz, $D_0$= 35.0567(17) kHz, $H_0$=0 (fixed value);
$\sigma$=14 kHz.\\ 
}
\label{Table2}
\end{table}

We first examined the chemistry by assessing the $J = 9.5 \leftarrow 8.5$ transition 
($e$ parity) at 438050.1 MHz of the NS radical ($^2\Pi_{1/2}$) in a positive column discharge.
Compared to H$_2$S, CS$_2$ was more efficient and hold stable discharge conditions.
Contrary to expectation, NO yielded a S/N enhancement of 2-3 over N$_2$
and atmospheric air instead of NO further increased the S/N by a factor 2.

The astronomical line at 250481.477 MHz was then inspected. It was readily detected with
optimum conditions analogous to those used for producing NS, i.e. CS$_2$ 
and atmospheric air were preferable : P(argon) = P(air) = 7.5 mTorr, P(CS$_2$) = 3.5 mTorr.
We have afterwards measured the transition frequencies between $150-850$ GHz, corresponding
to $J' = 3 - 17$.
Our error measurement is better than 50~kHz except for the line measured near 
851~GHz ($\approx 120$~kHz).

All lines were recorded in a positive column discharge, using $V_\mathrm{dc} = 1.4$ kV and
$I_\mathrm{dc}=50$ mA (ballast resistor $\approx$ 10 k$\Omega$).
To prove that the carrier is a cation, the discharge conditions were switched to 
a magnetically confined negative glow discharge,
applying $V_\mathrm{dc} = 4.2$ kV and $I_\mathrm{dc}=20$ mA (ballast resistor 
$\approx$ 100 k$\Omega$).
Indeed, it is known that the negative glow (a nearly electric field-free region) is the region
that produces a far higher concentration of cations (compared to the positive column),
and it is a common technique to lengthen the negative glow by
applying a longitudinal magnetic field \citep{DeLucia1983}, as this region is 
usually very short (a few cm).
Enhancements in signal strength (up to two orders of magnitude) with
this approach are expected for close-shell cations only.

The $J = 7 \leftarrow 6$ transition was monitored while increasing the 
intensity of the magnetic field.
The line, very weak without magnetic field, was about 5 times as strong, compared
to the positive column discharge conditions, for a magnetic field of about 200 G.
The measured frequency of this transition was identical (within 5 kHz) 
in both discharge conditions,
showing that in the positive column lines are unshifted by the Doppler effect.
This is as expected given the mass of the cation.
Only lightest ions such as SH$^-$ \citep{Civis1998}, OD$^-$ and N$_2$H$^+$ \citep{Cazzoli05}
have their frequency Doppler-shifted in the microwave region.

Various laboratory lines are shown in Figure~1
and Table~2 provides laboratory frequencies and the derived rotational and distortion constants.
The agreement with the astrophysical and $ab$ $initio$ constants
is excellent. Since the cation contains nitrogen and sulfur we conclude
that the only possible carrier is NS$^+$.
The observed S/N was too weak to observe N$^{34}$S$^+$ in natural abundance.

\section{Discussion}

The chemistry of NS and NS$^+$ in dense interstellar clouds has been poorly investigated so far 
(e.g., \citealt{Agundez2013}). The chemistry of sulfur in molecular clouds has been recently 
revisited by \citet{Fuente2016} and 
\citet{Vidal2017}, although unfortunately no prediction was provided for the abundances of NS or 
NS$^+$ in these studies. Here we consider the chemistry of dense clouds, with special attention 
to NS and NS$^+$. Nitrogen sulfide (NS) is assumed to be formed through the neutral-neutral reactions
\begin{equation}
\rm N + SH \rightarrow NS + H,
\end{equation}
\begin{equation}
\rm S + NH \rightarrow NS + H,
\end{equation}
for which the adopted rate constants are 10$^{-10}$ cm$^3$ s$^{-1}$ (KIDA 
database; Wakelam et al. 2015), while it is mostly destroyed through reactions with neutral atoms
(rates in units of 10$^{-11}$ cm$^3$ s$^{-1}$ from KIDA database)
\begin{equation}
\rm NS + O \rightarrow NO + S, k=10 
\end{equation}
\begin{equation}
\rm NS + N \rightarrow N_2 + S, k=3
\end{equation}
\begin{equation}
\rm NS + C \rightarrow CN + S, k=15
\end{equation}
The resulting modeled abundance of NS is thus very dependent on the assumed 
elemental abundances of N, O, S, and on the C/O elemental ratio. The NS$^+$ ion, on the other hand, is principally formed through reactions of atomic nitrogen with the SO$^+$ and SH$^+$ molecular ions 
\begin{equation}
\rm N + SO^+ \rightarrow NS^+ + O, \label{Reaction_N_SO+}
\end{equation}
\begin{equation}
\rm N + SH^+ \rightarrow NS^+ + H, \label{Reaction_N_SH+}
\end{equation}
The rate coefficient of reaction~(\ref{Reaction_N_SO+}) is of the order of $5\times10^{-11}$ cm$^3$ s$^{-1}$, 
according to a flowing afterglow measurement by \citet{Fehsenfeld1973}. No experimental information is 
available for reaction~(\ref{Reaction_N_SH+}), which is estimated to be $7.4\times10^{-10}$ cm$^3$ s$^{-1}$ 
in the KIDA database. NS$^+$ is principally destroyed through dissociative recombination with electrons and 
reactions with neutral atoms (O, N, and C). No measurement nor theoretical estimate is yet available for the 
dissociative recombination ($DR$) of NS$^+$ and a conservative value of 2 $\times $ 10$^{-7}$  (T/300)$^{-0.5}$ cm$^3$ s$^{-1}$ 
is given in the KIDA database, thought to be representative for $DR$ of diatomic 
ions. It is worth noting that observations show a rather uniform NS/NS$^+$ column density ratio across the observed sources 
(see Table~\ref{Table1}) which is not directly linked to any formation scenario of NS$^+$ from NS. 
Indeed, the charge exchange reaction of NS with H$^+$ and C$^+$, although thought to be rapid, will not be an 
efficient source of NS$^+$ in dense clouds where atomic ions are very rare. 

We display in Figure \ref{Fig_chemistry} gas-phase time-dependent model results for NS and NS$^+$ as well as for NO, 
NO$^+$, SO, and SO$^+$ for dense cloud conditions typical of B1b \citep{Cernicharo2012b}.
The chemical network is built from that used in \cite{roueff:15} where the coupling between N and S 
chemistries involving NS, NS$^+$, HNS$^+$ has been introduced, following the reactions available in the 
KIDA database. C, N, and S are assumed to be depleted to account for adsorption on grains: C/H = $7\times10^{-6}$, 
N/H =  $1\times10^{-6}$, S/H = $8\times10^{-8}$, and C/O = 0.7. The fractional abundances of NS, 
NS$^+$, SO$^+$ and NO$^+$ ions in B1b are shown to be reasonably reproduced whereas the SO and 
NO neutral radicals are somewhat underproduced and overproduced, respectively, in 
the model. More experimental and theoretical efforts are obviously needed to better 
constrain the chemistry of NS and NS$^+$ in molecular clouds.

A large NS/NS$^+$ ratio, as observed in hot cores, is the signature of a different formation mechanism of NS$^+$
which is rather coming from charge exchange reactions of relatively abundant NS with C$^+$ and H$^+$.
Although the fraction of sulfur locked by NS$^+$ in the observed sources is relatively low 
(10 ppm at most), the fact that this ion is ubiquitous in dark clouds of different evolutionary 
stages makes it a good potential tracer of the physical and chemical conditions of such environments.

\acknowledgements

We thank funding support from Spanish MINECO (grants AYA2012-32032, 
AYA2016-75066-C2-1-P, CSD2009-00038, and RyC-2014-16277) and from European research Council 
(grant ERC-2013-SyG 610256, NANOCOSMOS). This work was also supported by the Programme National 
PCMI
CEA and CNES, and the French National Research Agency (ANR-13-BS05-0008-02 ''IMOLABS'').

\end{document}